\documentclass{ws-p8-50x6-00}
  \newcommand{\epsfaxhax}[2]{
          \centerline{
            \hspace{-15pt}
            \epsfxsize=160pt
            {\epsfbox{#1}}
            \hspace{-7pt}
            \epsfxsize=160pt
            {\epsfbox{#2}}}
  }

\begin{document}

\title{Diquark Condensation in Dense Matter \\ a Lattice Perspective}

\author{Simon Hands and Susan Morrison}

\address{Department of Physics, University of Wales Swansea, Singleton Park,\\
Swansea SA2 8PP, UK\\E-mail: s.hands,s.morrison@swansea.ac.uk}


\maketitle

\abstracts{
We review efforts to study, using the methods of lattice field theory,
the phenomenon of diquark condensation
via BCS pairing at a Fermi surface, which has been 
proposed as a mechanism for color superconductivity in dense quark matter.
The particular models studied are the Gross-Neveu
model and SU(2) lattice gauge theory; in both cases evidence for 
superfluidity at high density is presented. The behaviour expected for
quarks in both fundamental and adjoint representations of SU(2) is contrasted. 
}

\section{Introduction}
At low temperatures, fermionic matter becomes degenerate; ie. 
all single particle states are occupied up to some sharply defined scale, 
the {\sl Fermi energy\/}, which characterises allowed physical processes.
In a normal Fermi liquid, excitations above the Fermi energy can have 
arbitrarily small energy. However, if there is an attractive interaction between
fermion pairs, and the Fermi energy is large enough, then the normal state
should be 
unstable with respect to the opening of a gap at the Fermi surface. In 
field-theoretic language, this gap is signalled by a non-vanishing
condensate formed from two elementary fermion fields. If our fermions are
quarks, then we have a {\sl diquark condensate\/}
$\langle qq\rangle\not=0$. As reviewed below, diquark condensation
can lead to new and more interesting ground states.
In this talk we will summarise our efforts in, and
future prospects for, probing this phenomenon using the methods of lattice field
theory. Throughout we will be attempting to explore the zero temperature
limit, although this may be difficult to reach in practice on finite volumes.

When considering the possible diquark condensates that may form, 
various questions
present themselves. Firstly, if the quarks carry a gauge charge, 
is the condensate gauge invariant? If so, then 
the state with $\langle qq\rangle\not=0$ breaks some global symmetry of
the Lagrangian, implying eg. that excitations carry indefinite baryon number.
This is the phenomenon of {\sl superfluidity\/}, which occurs in condensed
matter physics at milliKelvin temperatures for $\mbox{He}^3$. The systems
we have studied to date are expected to exhibit superfluidity. However,
if the condensate breaks a local symmetry, then the system will be a
{\sl superconductor\/}, and the diquark condensate a direct analogue of 
the Cooper pairing found in metals at low temperature. Since the quarks of 
QCD transform under a non-Abelian symmetry, the phenomenon is still more
exotic: {\sl color superconductivity\/}
has been postulated as a dynamical embodiment
of the Higgs mechanism, making some or all of the gluons massive at high 
baryon density.\cite{BL,ARW,RSSV} Secondly, 
is the condensate a spacetime scalar?
Naively one might expect so, but examples of rotationally
non-invariant condensates
are known in condensed matter physics ($\mbox{He}^3\mbox{A}$), 
and have been postulated in QCD.\cite{ARW} A parity violating diquark 
condensate is another possibility.

The most crucial consideration is that the condensate must respect 
the Pauli Exclusion Principle, so that its wavefunction is antisymmetric
under exchange of all available quantum numbers, which in the case
of quarks are spacetime, color and flavor. This has the consequence that the
ground state may be extremely sensitive to the number of light flavors 
present,\cite{ARW} or as we shall see below, to the representation of the 
color group carried by the quarks. This impact of the Exclusion Principle
makes the phenomenon of intrinsic theoretical interest, and may even underlie
the known difficulties in simulating QCD with non-zero baryon
density.\cite{IMB}

\section{Four Fermi Models}

The first model we will consider is referred to as either the ``Gross-Neveu''
(GN) model in 2+1, or the ``Nambu -- Jona-Lasinio'' (NJL) model in 3+1
dimensions, and is a relativistic generalisation of the BCS model originally
used to describe superconductivity. It
has the following Lagrangian density in continuum notation:
\begin{equation}
{\cal L}=\bar\psi(\partial{\!\!\! /}\,+m)\psi-g^2\left[
(\bar\psi\psi)^2-(\bar\psi\gamma_5\vec\tau\psi)^2\right],
\end{equation}
where $\psi$ is an isodoublet transforming under a global 
axial $\mbox{SU(2)}_L\otimes\mbox{SU(2)}_R$ symmetry, which is exact in the 
chiral limit $m\to0$:
\begin{equation}
\psi_L\mapsto U\psi_L\,,\,\bar\psi_L\mapsto\bar\psi_L U^\dagger\;\;;\;\;
\psi_R\mapsto V\psi_R\,,\,\bar\psi_R\mapsto\bar\psi_R V^\dagger.
\label{eq:SU22}
\end{equation}
${\cal L}$ is also invariant under a global $\mbox{U(1)}_V$ of baryon number:
\begin{equation}
\psi\mapsto e^{i\alpha}\psi\;\;;\;\;\bar\psi\mapsto\bar\psi e^{-i\alpha}.
\label{eq:U1B}
\end{equation}

Let us review some properties of the model. First, for sufficiently strong 
coupling $g^2$, the axial symmetry
spontaneously
breaks to a $\mbox{SU(2)}_V$ of isospin by the formation of a chiral condensate
$\langle\bar\psi\psi\rangle$. The spectrum in the chirally broken phase 
contains massive ``baryons'' -- the elementary fermions, and $\psi\bar\psi$
composite ``mesons'', which include three light Goldstone pions. The 
chiral condensate sets the dynamical mass scale $M/\Lambda$ for the model, 
which can thus be taken to zero at some critical $g_c^2$, defining a continuum
limit. A special feature of the GN model is that the continuum limit defines
an interacting field theory, whereas that of the NJL model is logarithmically
trivial.\cite{HKK} For the present purposes, the most interesting feature
of the models is that they can be formulated on a lattice, and simulated
with baryon chemical potential $\mu\not=0$ by standard Monte Carlo
methods;\cite{GNsim} 
the reasons for this are subtle and only recently understood.\cite{BHKLM}
It is found that for $\mu$ greater than some $\mu_c$ of order the
baryon mass the symmetry (\ref{eq:SU22}) is restored.

In our work\cite{HM} 
we have focussed on the formation at large $\mu$ (for non-interacting
fermions $\mu$ is precisely the Fermi energy) of a diquark condensate 
$\langle qq\rangle$,
where in continuum notation the diquark wavefunction in 2+1 dimensions reads
\begin{equation}
qq=\psi^{tr}({\cal C}\gamma_5)\otimes\tau_2\otimes\tau_2\psi.
\end{equation}
Here ${\cal C}$ is the charge conjugation matrix, and the ${\cal C}\gamma_5$
structure ensures that the condensate is a scalar. The first of the $\tau_2$
matrices operates on an implict flavor space due to the species doubling 
generic to lattice fermions, and the second on the explicit isospin indices.
Since all three operators are antisymmetric matrices, the overall antisymmetry 
of the wavefunction is ensured. This condensate respects the axial 
symmetry (\ref{eq:SU22}) but spontaneously breaks $\mbox{U(1)}_V$
(\ref{eq:U1B}).

\subsection{Two-Point Function Approach}
Our first attempt to look for a signal for $\langle qq\rangle\not=0$ was via
the asymptotic behaviour of the diquark propagator, which by the cluster
property should be proportional to the square of the condensate:
\begin{equation}
G(x)=\langle qq(0)\bar q\bar q(x)\rangle=\langle qq(0)\bar q\bar q(x)\rangle_c
+\langle qq\rangle\langle\bar q\bar q\rangle
\Rightarrow\lim_{x\to\infty}G(x)=\vert\langle qq\rangle\vert^2.
\label{eq:cluster}
\end{equation}
The condensate should thus manifest itself as a plateau in the large-$t$
behaviour of the timeslice propagator. Results\cite{HM} 
for $G(t)=\sum_{\vec x}G(\vec x,t)$ 
from a $16^2\times40$ lattice,
with $\mu=0.8>\mu_c$, are shown in Fig. 1, clearly showing a stable plateau
when compared to the same quantity
obtained for non-interacting fermions, shown with closed
symbols. Taking the square root of the plateau height as a measure for 
$\langle qq\rangle$ results in the plot shown in Fig. 2, where chiral symmetry
restoration at $\mu_c\simeq0.65$ followed by a rapid rise in baryon number
density is also clearly visible.
\begin{figure}[t]
\epsfaxhax{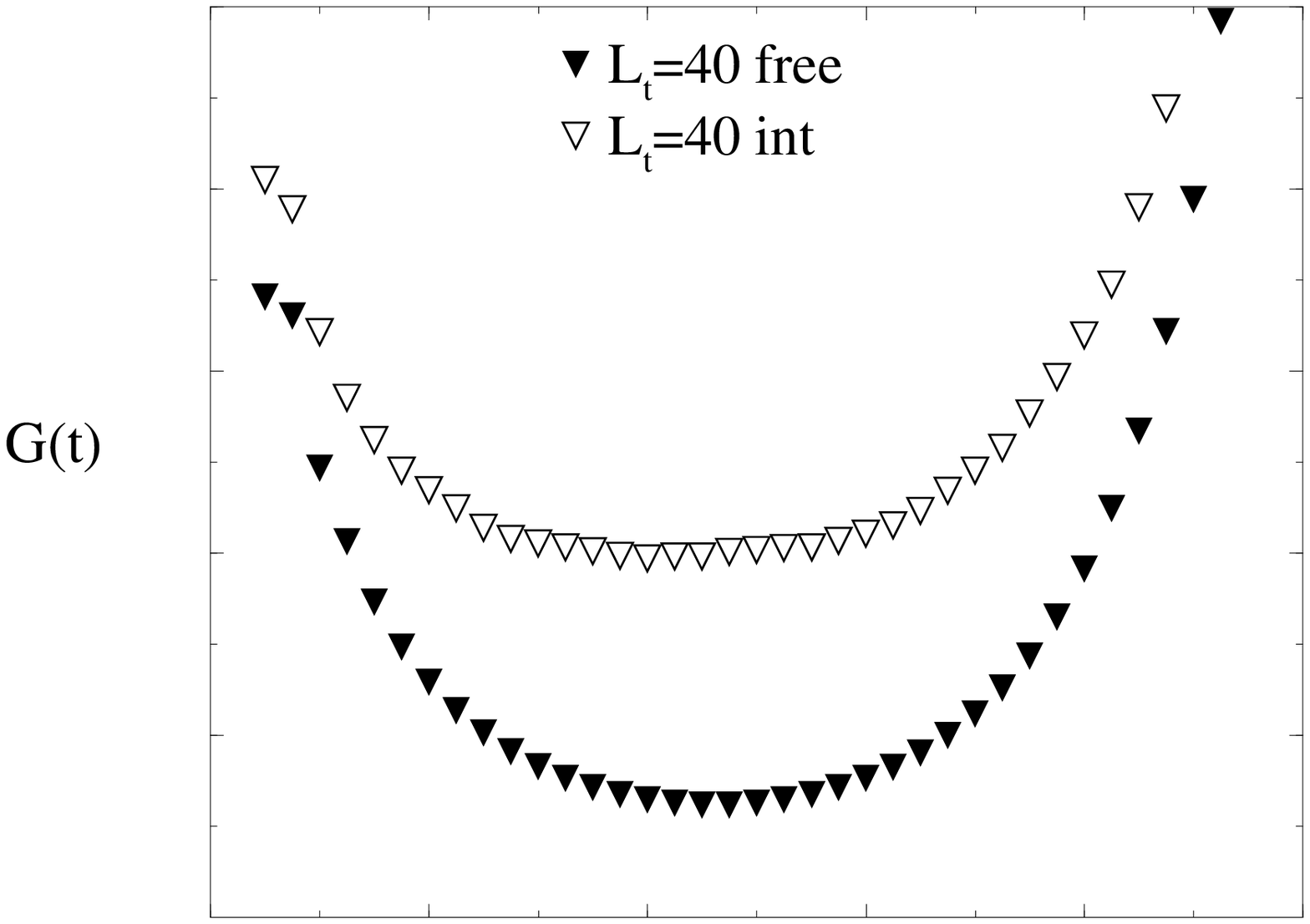}{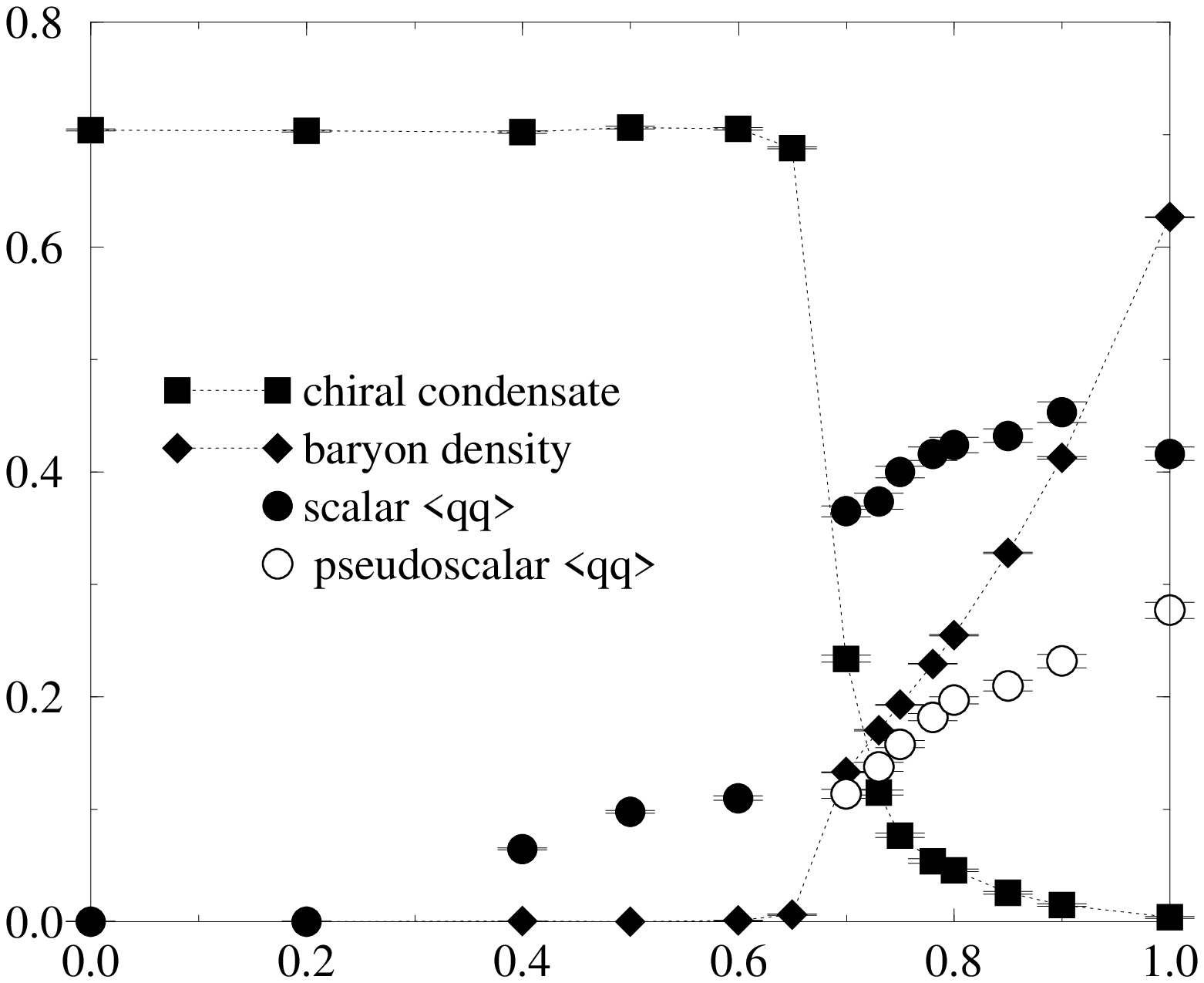}
\vskip -0.3cm
\caption{\hskip 4.1cm Figure 2.}
\vskip -0.3cm
\setcounter{figure}{2}
\end{figure}
We see a dramatic increase in $\langle qq\rangle$ proceeding from 
the low density chirally broken phase to the high density chirally symmetric
phase, suggesting that diquark condensation is taking place 
across the transition, as revealed by long range order
in the timelike direction. This result is consistent with the notion 
of competition between chiral and diquark condensates.\cite{ARW}
We have also observed signals both for a small scalar
diquark condensate in the broken phase, and for a non-vanishing 
{\sl pseudoscalar\/}
condensate in the dense phase, implying spontaneous breaking of parity; both 
these may well be finite volume
artifacts, however.

Unfortunately, data from lattices with spatial volume $L_s^2$
varying from $8^2$
to $24^2$ do not support a naive application of Eq. (\ref{eq:cluster}), which 
suggests the plateau height should be an extensive quantity. Indeed, 
the plateau height saturates for $L_s\geq16$, a result which may be due to the
influence of Goldstone fluctuations, which would wash out the signal in the
absence of an external source, or from the relatively small number of
participating $qq$
states in the vicinity of the Fermi surface on these moderate systems.

\subsection{One-Point Function Approach}

In an attempt to clarify matters we have also performed direct 
measurements of the diquark one-point function $\langle qq\rangle$.
By rewriting the fermionic action $S_{ferm}=\bar\psi M\psi$ in the 
{\sl Gor'kov representation\/}, it is possible to add explicit diquark 
source terms:
\begin{eqnarray}
S_{ferm}&=&(\bar\psi,\psi^{tr})\left(\matrix{\bar\jmath\tau_2&{1\over2}M\cr
-{1\over2}M^{tr}&j\tau_2\cr}\right)\left(\matrix{\bar\psi^{tr}\cr\psi\cr}
\right)\equiv\Psi^{tr}{\cal A}[j,\bar\jmath]\Psi;\nonumber\\
Z[j,\bar\jmath]&=&\langle\mbox{Pf}{\cal A}[j,\bar\jmath]\rangle.
\end{eqnarray}
The diquark condensate is then defined by
\begin{equation}
\langle qq\rangle={1\over V}{{\partial\ln Z}\over{\partial j}}\biggr\vert_
{j,\bar\jmath=0}=\displaystyle\lim_{j,\bar\jmath\to0}{1\over V}
\biggl\langle{1\over2}\mbox{tr}\left\{{\cal A}^{-1}\left(
\matrix{&\cr&\tau_2\cr}\right)\right\}\biggr\rangle,
\label{eq:qq(j)}
\end{equation}
which is straightforward to implement. Our results for
$\langle qq(j)\rangle$ in the GN model from a $16^2\times24$ lattice are shown 
in Fig. 3.
\begin{figure}[t]
\epsfaxhax{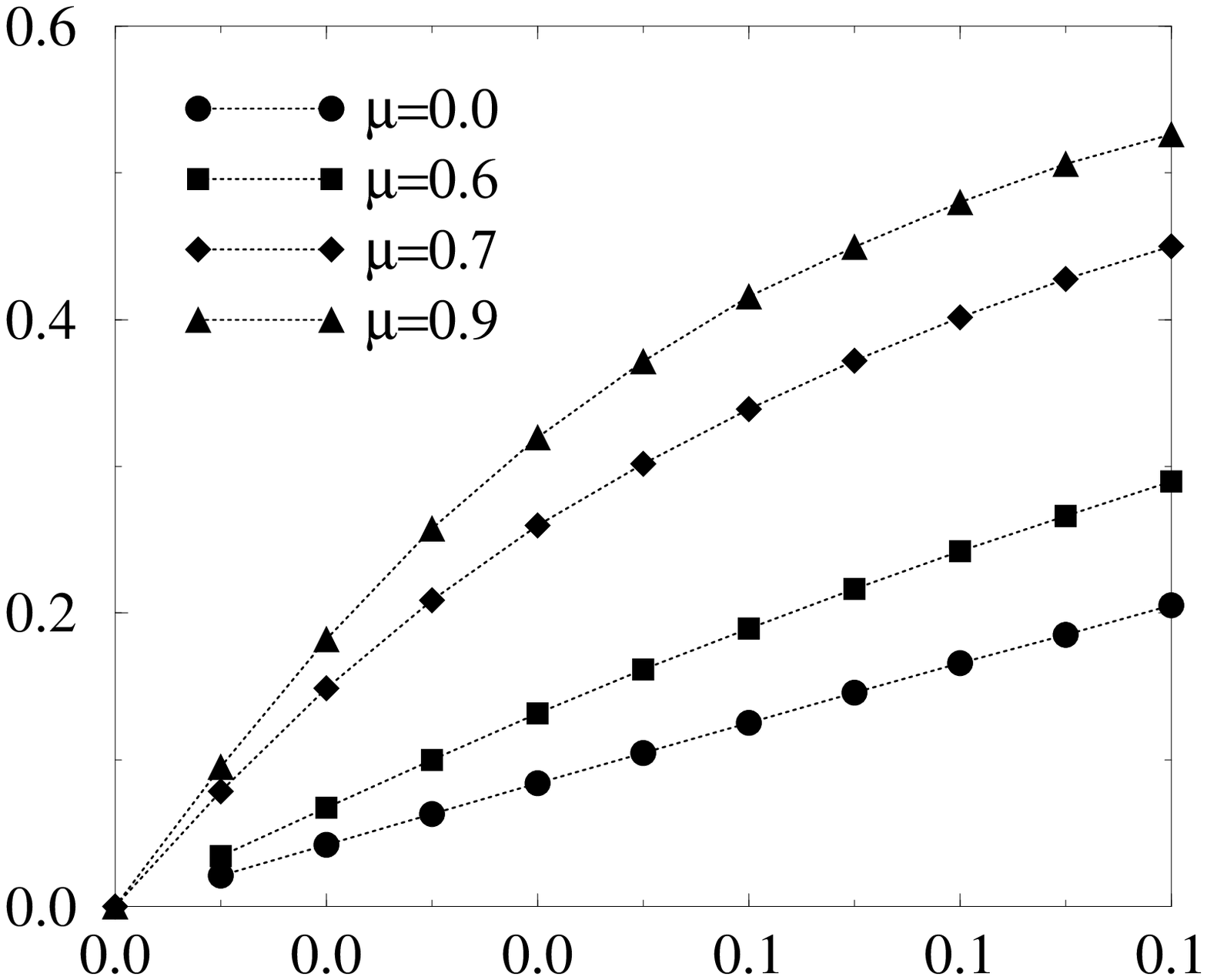}{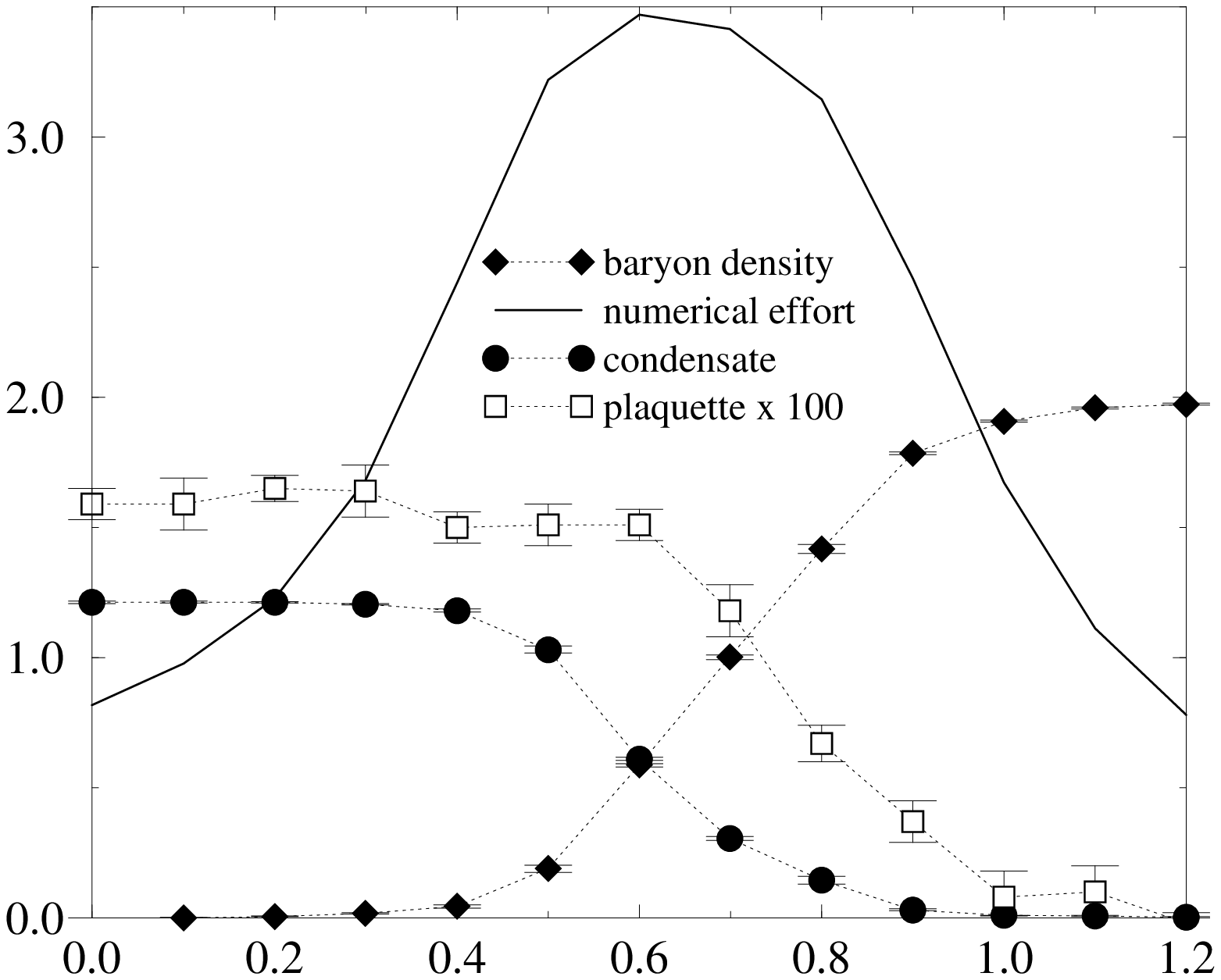}
\vskip -0.3cm
\caption{\hskip 4.1cm Figure 4.}
\vskip -0.3cm
\setcounter{figure}{4}
\end{figure}
We see a large jump in the signal as the chiral transition takes place; at low
density the signal is approximately linear in $j$, whereas at high density
there is a marked curvature. In neither case, however, is there convincing 
evidence for a non-zero intercept as $j\to0$.

It should be stressed that these measurements are ``quenched'' in the
sense that the diquark source term is
not included in the Monte Carlo update
algorithm. We are currently attempting full simulations with $j\not=0$ in 
the expectation that finite
volume effects will be easier to interpret.\cite{HLM} 
Inclusion of the source term will also enable estimates of disconnected
contributions to the two-point function, which may have a better
overlap with the Goldstone mode if one exists. 
It will also be valuable to develop
spectroscopy techniques in the Gor'kov representation, since in the presence of
a gap the eigenstates of the transfer matrix will be a superposition of $q$ 
and $\bar q$ states. It is safe to say that much more work is needed before
the apparently straightforward 
exercise presented by four fermi models is complete.

\section{Two Colors QCD}

\subsection{Fundamental Quarks}

SU(2) gauge theory with staggered lattice fermions in the fundamental
representation 
does not suffer from the difficulties 
associated with simulating dense QCD, since it can be shown
that $\mbox{det}M$ is real\footnote{This follows 
readily from $\mbox{det}M=\mbox{det}\tau_2M\tau_2=\mbox{det}M^*$.}
and positive for all $\mu$\cite{BO}. 
In Fig. 4 we show results from runs on a $4^4$ lattice, with $N_f=4$ physical
flavors of fermion and $m=0.2$, in the strong gauge coupling limit as a 
function of $\mu$.
For $\mu<0.4$, the system remains essentially unchanged, but for 
$0.4<\mu<0.9$ the chiral condensate
$\langle\bar\psi\psi\rangle$ 
smoothly decreases to zero, at the same time as the 
baryon density $n\equiv\langle\bar\psi\gamma_0\psi\rangle$ increases from zero 
to its saturation value of two quarks per lattice site. The numerical effort
required, as measured by the number of computer iterations required
to invert $M$, rises steeply in the crossover region. Most interestingly,
the average plaquette decreases as density rises, until at saturation it 
assumes the expected quenched value of zero. This is a signal of {\sl Pauli
blocking\/}; at high density virtual $q\bar q$ pairs are suppressed for
kinematical reasons, and color screening via vacuum polarisation thus reduced.

A physically appealing way of understanding the role of the complex
phase of $\mbox{det}M$ in gauge theories 
is in terms of {\sl conjugate quarks\/}.\cite{GGS}
Monte Carlo simulations of QCD demand a real functional measure, and therefore
update using $\mbox{det}MM^\dagger=\mbox{det}M\mbox{det}M^*$. The $M$
describes quarks $q$ in the {\bf 3} representation of SU(3), 
the $M^*$ conjugate 
quarks $q^c$ in the ${\bf\bar3}$. Usually conjugate quarks are regarded as 
anti-quarks of a different flavor, but once $\mu\not=0$ it can be checked 
that both $q$ and $q^c$ couple to $\mu$ in the same way and hence
carry positive baryon number. Therefore simulations with a real measure
can describe gauge singlet $qq^c$ bound states; in particular
a pseudo-Goldstone state, the {\sl baryonic pion\/} can form,\cite{LKS} 
and is the lightest 
baryon in the spectrum, since by the usual PCAC arguments its mass is
expected to vanish as $\surd{m}$ in the chiral limit. Now, simple energetic
arguments suggest that observables should start to show $\mu$-dependence
for $\mu\geq\mu_o\simeq m_{lb}$, where $m_{lb}$ is the lightest baryon mass.
The presence of an unphysical light baryon therefore plays havoc with the 
physics of $\mu\not=0$.\footnote{For four fermi models, $q\bar q$ states
are much lighter than $qq^c$ due to disconnected diagrams.\cite{BHKLM}}

For SU(2) gauge theory, however, baryonic pions are not unphysical. Consider
the kinetic term for staggered lattice fermions:
\begin{equation}
S_{kin}={1\over2}\sum_{x,\mu=0,3}\eta_\mu(x)
\left[\bar\chi(x)U_\mu(x)\chi(x+\hat\mu)-\bar\chi(x)U_\mu^\dagger(x-\hat\mu)
\chi(x-\hat\mu)\right].
\end{equation}
We can identify both $\mbox{U}(1)_V$ and $\mbox{U}(1)_A$ global 
symmetries, which may be associated with baryon number and axial charge 
conservation respectively:
\begin{equation}
\mbox{U}(1)_V:\;\chi\mapsto e^{i\alpha}\chi\;\;\bar\chi\mapsto\bar\chi
e^{-i\alpha}\;\;\;\;\mbox{U}(1)_A:\;\chi\mapsto e^{i\beta\varepsilon}\chi\;\;
\bar\chi\mapsto\bar\chi e^{i\beta\varepsilon},
\label{eq:U1VA}
\end{equation}
where $\varepsilon(x)=(-1)^{x_0+x_1+x_2+x_3}$. However, by defining new fields
\begin{equation}
\bar X_e=(\bar\chi_e,-\chi_e^{tr}\tau_2),\;\;\;X_o^{tr}=
(\chi_o^{tr},\bar\chi_o\tau_2),
\end{equation}
where the subscripts distinguish between even and odd lattice sites, then 
due to the pseudoreality of the {\bf2} representation the 
action may be recast as
\begin{equation}
S_{kin}={1\over2}\sum_{x even, \mu}\eta_\mu(x)\left[
\bar X_e(x)U_\mu(x)X_o(x+\hat\mu)-\bar X_e(x)U_\mu^\dagger(x-\hat\mu)
X_o(x-\hat\mu)\right].
\label{eq:Skin}
\end{equation}

Since the $X$ and $\bar X$ fields have two components, the global symmetry is 
enlarged to U(2), and can be viewed as relating mesonic
$q\bar q$ states to baryonic $qq$ states. In particular, a condensate
$\langle\bar\chi\chi\rangle$, which in the chiral limit at zero
density would be expected to break spontaneously 
the original $\mbox{U}(1)_A$ symmetry, can be U(2)-rotated into a 
gauge invariant diquark condensate 
$\langle qq_{\bf2}\rangle\equiv\langle\chi^{tr}\tau_2\chi\rangle$,
which also breaks $\mbox{U}(1)_V$. In either case the overall symmetry
breaking is U(2)$\to$U(1), which is different from the pattern 
$\mbox{SU}(2N_f)\to\mbox{Sp}(2N_f)$ predicted for continuum SU(2) gauge 
theory.\cite{Peskin} A full analysis\cite{HKLM} shows that in the limit 
$m\to0$, $\mu\to0$ the symmetry breaking is accompanied by three Goldstone
modes, with quantum numbers $0^-$, $0^+$, $0^+$. Depending on the direction
on the U(2) manifold chosen by the condensate, the Goldstones can be regarded
as either mesons or baryons, and for this reason the presence of baryonic
pions in the Monte Carlo simulation is not harmful.\footnote{ 
For the model described by the measure $\mbox{det}MM^\dagger$ the 
actual pattern is U(4)$\to$O(4).}

\begin{figure}[t]
\epsfaxhax{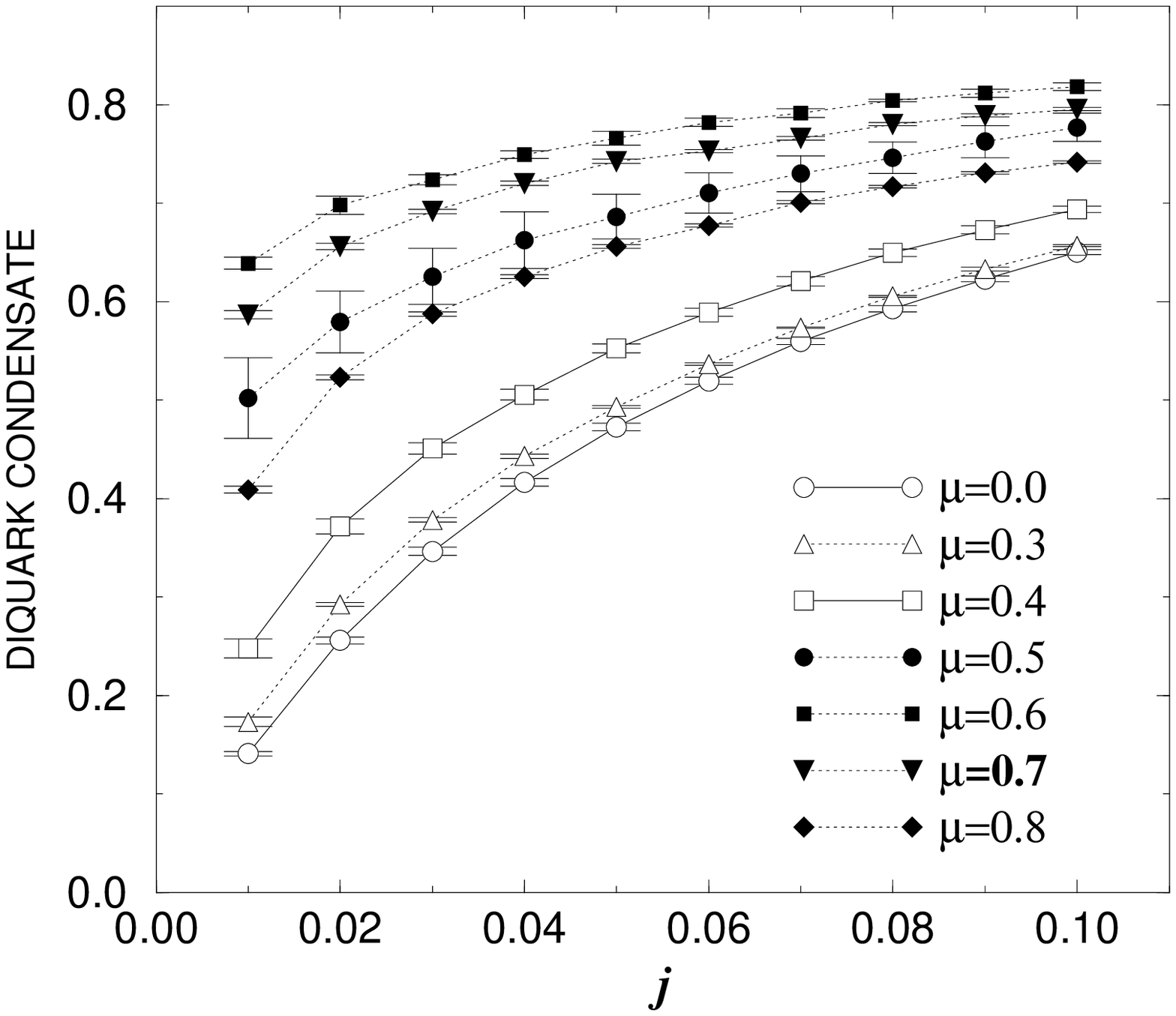}{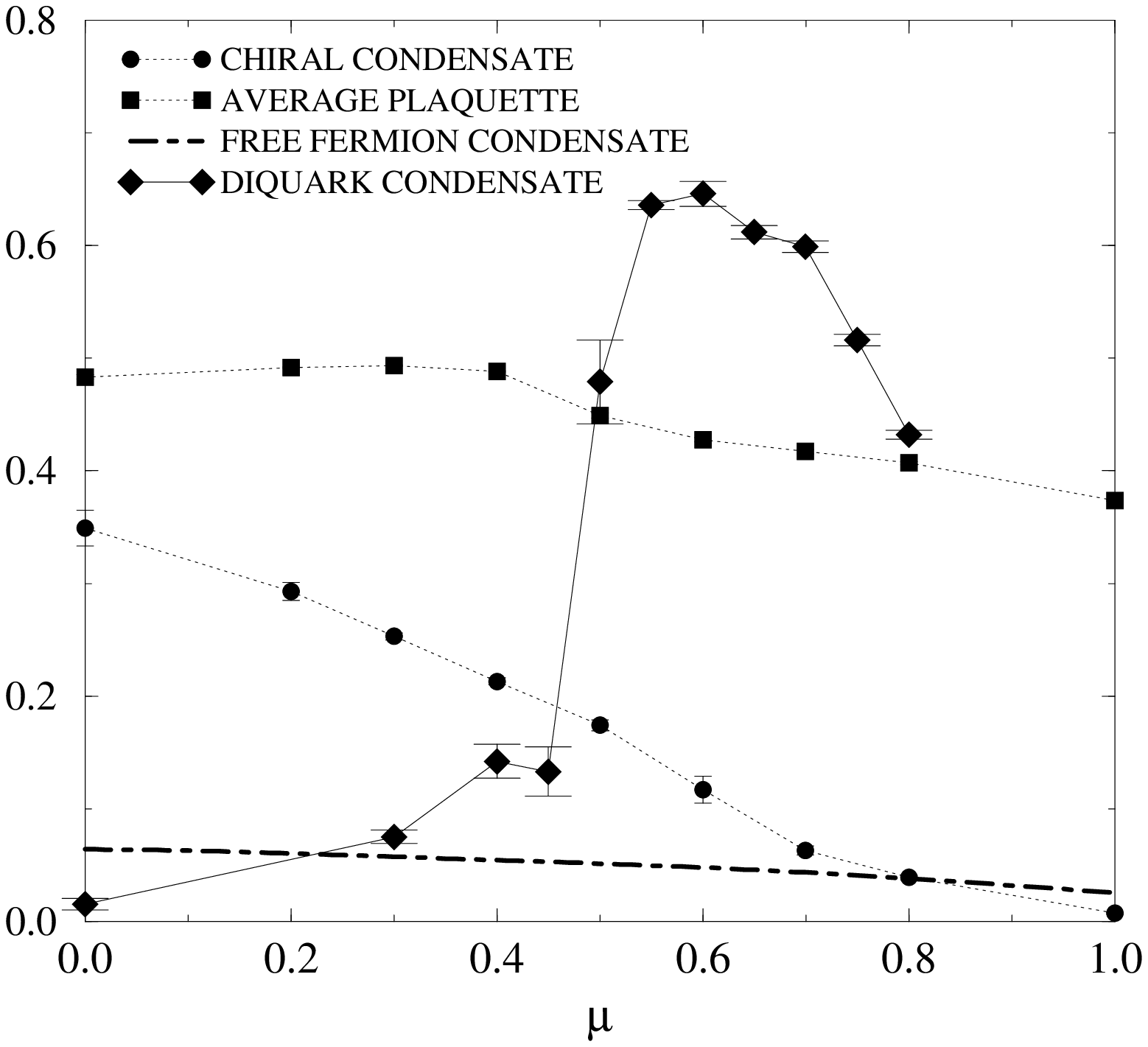}
\vskip -0.3cm
\caption{\hskip 4.1cm Figure 6.}
\vskip -0.3cm
\setcounter{figure}{6}
\end{figure}
Once $m$ and $\mu$ differ from zero, the high degree of symmetry is no longer
present. Let us first present some simulation results obtained using a
hybrid Monte Carlo (HMC) 
algorithm, which corresponds to $N_f=8$ physical flavors.
In Fig. 5 we show results for $\langle qq_{\bf2}(j)\rangle$ obtained using 
eq. (\ref{eq:qq(j)}) for various $\mu$ on a 
$6^4$ lattice with quark mass $m=0.05$, and gauge coupling $\beta=1.5$.\cite{MH}
There is a clear distinction between the curves for $\mu\leq0.4$, which appear
to extrapolate to zero as $j\to0$, and those for $\mu\geq0.5$, which plausibly 
yield $\langle qq\rangle\not=0$ in this limit. Results for $\langle
qq_{\bf2}\rangle$ obtained using a simple polynomial extrapolation to the 
zero source limit are plotted as a function of $\mu$ in Fig. 6, along with the
chiral condensate $\langle\bar\chi\chi\rangle$ and the average plaquette.
The chiral condensate appears to decrease smoothly as soon as $\mu>0$, until
by $\mu=0.8$ it actually falls below its free-field value. The immediate
onset of the fall with $\mu$ 
is difficult to understand unless there are thermal effects
associated with the rather small lattice volume. The diquark condensate, by 
contrast, stays small (probably even zero within systematic errors) until 
a value $\mu_c\simeq0.4$, whereupon it rises sharply to a large non-zero value,
and then almost immediately starts to fall again. Evidence for a phase 
transition at $\mu=\mu_c$ also comes from the average plaquette, which shows a
discernible kink between the low density phase, where it is approximately 
constant, and the high density phase where it decreases with $\mu$, 
presumably as a consequence of Pauli blocking.

It is tempting to associate
the decrease in $\langle qq_{\bf2}\rangle$ with $\mu$ in the dense phase
with asymptotic freedom, since the scale defined by the Fermi
energy rises monotonically with $\mu$, and further measurements reveal 
that $n$ 
also rises significantly over this range of $\mu$. 
It will require much more extensive
simulation on a range of lattice volumes and spacings, however, before this
effect can be distinguished from lattice artifacts due to saturation.

\begin{figure}[t]
\epsfaxhax{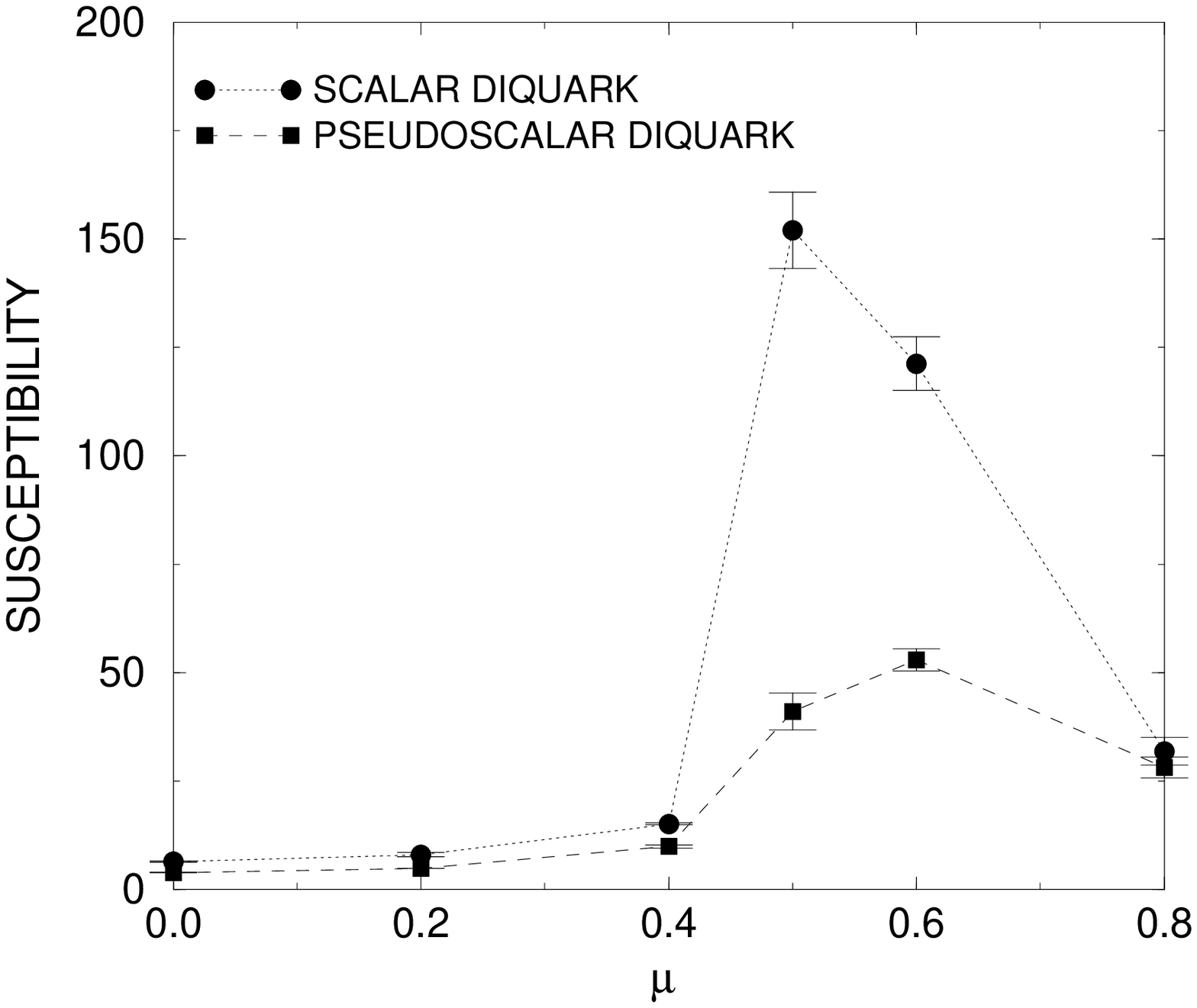}{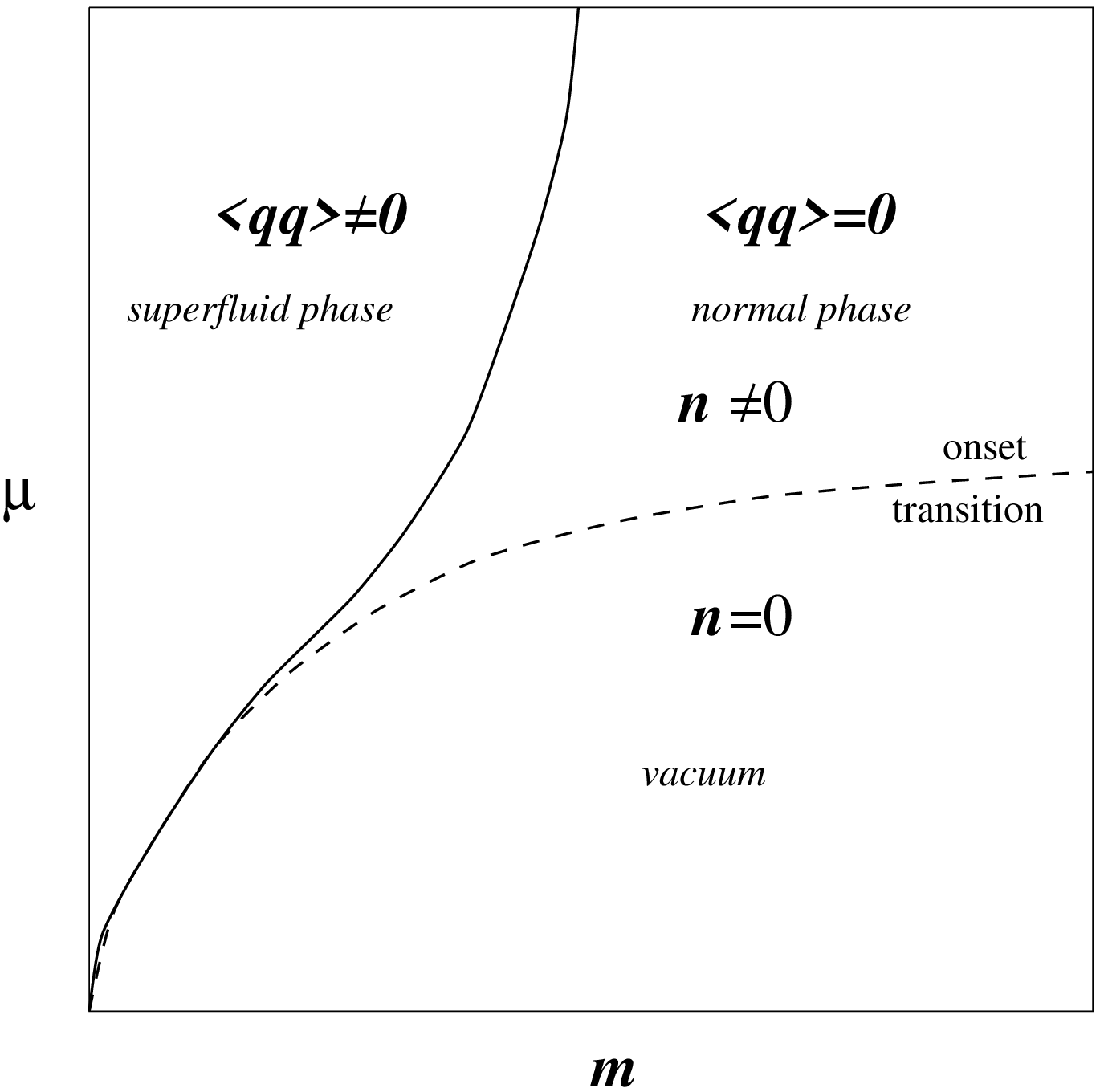}
\vskip -0.3cm
\caption{\hskip 4.1cm Figure 8.}
\vskip -0.3cm
\setcounter{figure}{8}
\end{figure}
As mentioned above, the diquark condensate spontaneously breaks the
original $\mbox{U}(1)_V$ symmetry of (\ref{eq:U1VA}), which remains a 
symmetry of the Lagrangian even once $m,\mu\not=0$. Therefore we expect the
appearance of the condensate in the dense phase to be accompanied by a
scalar Goldstone mode. In Fig. 7 we plot the contributions to the
diquark susceptibilities,
\begin{equation}
\chi_{qq}=\sum_x\langle qq(0)\bar q\bar q(x)\rangle,
\end{equation}
from connected quark line diagrams in both scalar and pseudoscalar channels.
Both increase sharply for $\mu>\mu_c$; the scalar signal, which should project
onto an exact Goldstone mode, exceeds the pseudoscalar, which projects onto 
a pseudo-Goldstone mode. The two channels should become degenerate in the 
chiral limit.\cite{HKLM} Once again, it is likely that to establish the level
ordering in the dense phase 
unambiguously will require the contribution of disconnected diagrams to
be taken into account.\cite{HKMS}

Fig. 8 is a suggested phase diagram for SU(2) lattice gauge theory with 
fundamental quarks in the $(m,\mu)$ plane. In principle there are three phases:
the {\sl vacuum\/}, 
for which both baryon number density $n$ and diquark condensate
$\langle qq\rangle$ vanish; a {\sl normal\/} phase for which $n>0$ but 
$\langle qq\rangle=0$; and a {\sl superfluid\/} phase\footnote{Unlike 
$\mbox{He}^3$, superfluidity arises here via
BCS pairing in the $s$-wave channel.} (recall $qq_{\bf2}$ is
gauge invariant)
for which both $n>0$ and the order parameter 
$\langle qq\rangle\not=0$. Note that the chiral condensate
$\langle\bar\chi\chi\rangle$, which is usually considered as an order 
parameter, is everywhere non-zero in this plane, although from Fig. 6 we expect 
it to decrease from bottom right to top left.
The dashed line separating the vacuum from the normal
phase is $\mu_o(m)\propto\surd m$, following the arguments about the lightest
baryon given above. The solid line $\mu_c(m)$ which separates the superfluid 
from the normal phase 
may well coincide with $\mu_o$ for some or all of its extent. It will 
be a goal of future simulations\cite{HKMS} to establish in the first instance
whether the normal phase exists, and if so whether it 
is confined to the 
large $m,\mu$ region where $\langle\bar\chi\chi\rangle$ is favoured
kinematically, but $\langle qq\rangle$ 
suppressed by asymptotic freedom, or extends in a narrow tongue all the
way to the zero density chiral limit at the origin. Another issue to explore
will be the persistence of the superfluid phase for $T>0$. Finally, even though
the model has the wrong physics for $\mu<\mu_c$, studies of gluonic dynamics in
the dense phase may still be of relevance for QCD;\cite{MPL} 
in particular we might expect competition between the enhanced screening
expected from a non-zero density of color sources in the ground state, and the 
kinematic anti-screening due to Pauli blocking.

\subsection{Adjoint Quarks}

The consequences of formulating the theory with adjoint rather
than fundamental quarks are profound. Even at zero density, the adjoint model
is distinct because of the possibility of gauge invariant spin-${1\over2}$ bound
states, either $qg$ or $qqq$, in the spectrum. We wish to advocate SU(2)
lattice gauge theory with adjoint quarks as a ``Toy QCD'' for the purposes of
non-zero density studies.

First let us discuss the symmetries of the kinetic term for
a single staggered flavor. The manipulations
leading to eq. (\ref{eq:Skin}) go through as before, but now the $U_\mu$ can 
be chosen real, and the $X,\bar X$ fields defined:
\begin{equation}
\bar X_e=(\bar\chi_e,\chi_e^{tr}),\;\;\;X_o^{tr}=(\chi_o^{tr},\bar\chi_o).
\end{equation}
Formation of a chiral condensate $\langle\bar\chi\chi\rangle$ causes a 
$\mbox{U}(2)\to\mbox{Sp}(2)$ symmetry breaking, resulting in one broken 
generator whose Goldstone 
can be identified with the familiar pion. Once again, this is
the opposite of the continuum result.\cite{Peskin} Since there are no light
diquark states in this case we expect no early onset, ie.
$\lim_{m\to0}\mu_o\not=0$.

Now consider possible diquark condensates which might form at high density.
In the absence of a detailed dynamical argument, we may enunciate three
plausible 
conditions for the $qq$ operator in the ``maximally attractive channel'':
\begin{itemize}

\item $qq$ is gauge invariant

\item $qq$ is a spacetime scalar

\item $qq$ is as local as possible in the lattice $\chi$ fields

\end{itemize}
The $qq_{\bf2}$ operator discussed in the previous subsection satisfies each
of these conditions. For adjoint quarks, however, the Exclusion Principle
dictates that one of the conditions must be violated, since the 
equivalent $qq_{\bf3}\equiv\chi^{tr}(x)\chi(x)$ vanishes identically.
A possible non-local chirally symmetric condensate could form from an operator
\begin{equation}
qq_{\bf3}^\prime=\sum_{\pm\mu}\eta_\mu(x)(-1)^{x_\mu}
\left[\chi^{tr}(x)U_\mu(x)\chi(x+\hat\mu)-\bar\chi(x)U_\mu(x)
\bar\chi^{tr}(x+\hat\mu)\right].
\end{equation}
This breaks $\mbox{U}(2)\to\mbox{U}(1)\otimes\mbox{U}(1)_A$, resulting in 
two $0^+$ Goldstones, one of which persists even once $m,\mu\not=0$. The
original $\mbox{U}(1)_V$ of (\ref{eq:U1VA}) is broken. A possible
phase diagram resulting from $\langle qq_{\bf3}^\prime\rangle$ condensation is 
shown in Fig. 9. One interesting issue is that although $qq_{\bf3}^\prime$ is
a scalar under a lattice version of parity,\cite{HKLM} once a transformation
is made to fermion fields with continuum degrees of freedom, it is a 
spacetime pseudoscalar.

\begin{figure}[t]
\epsfaxhax{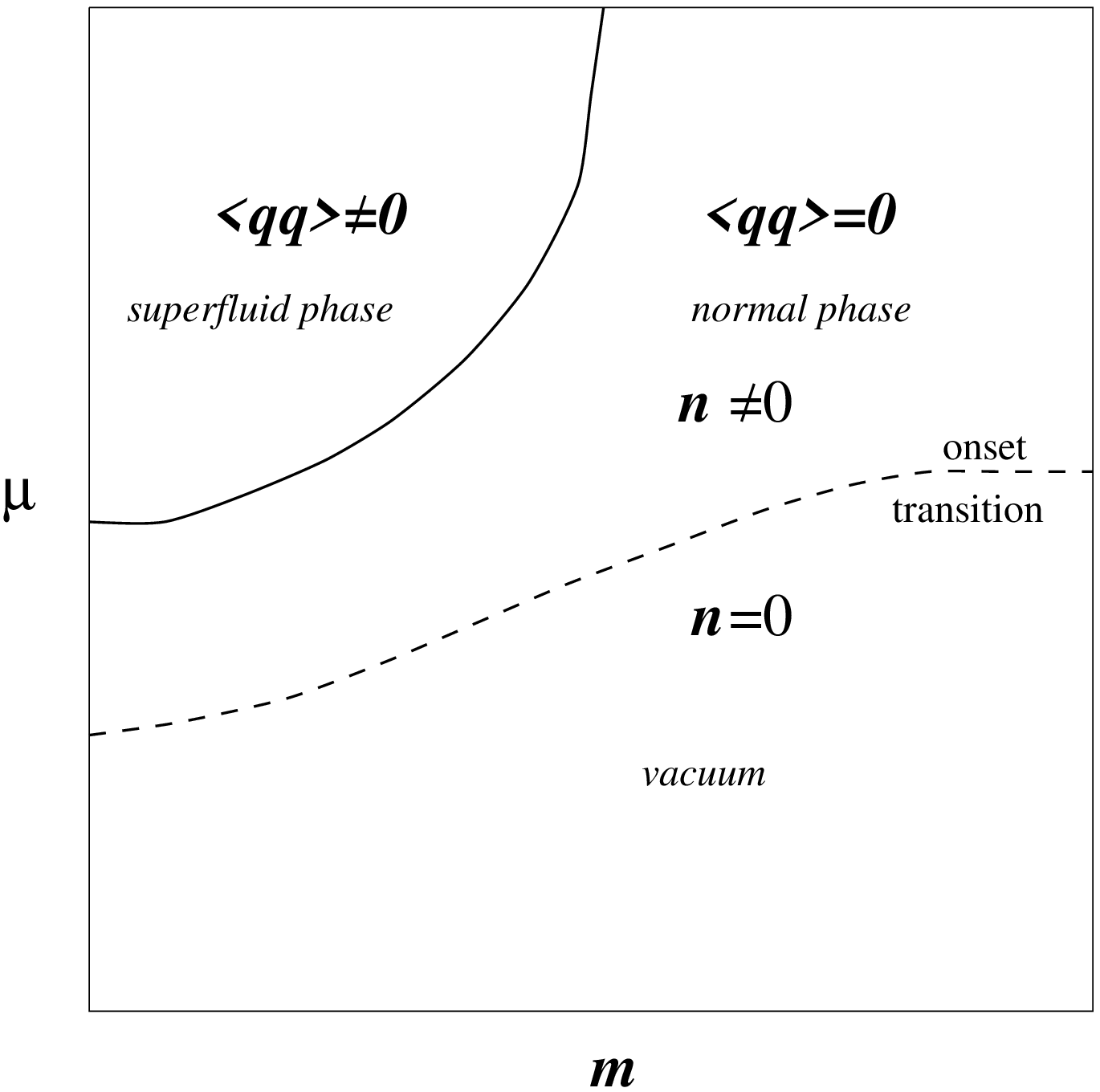}{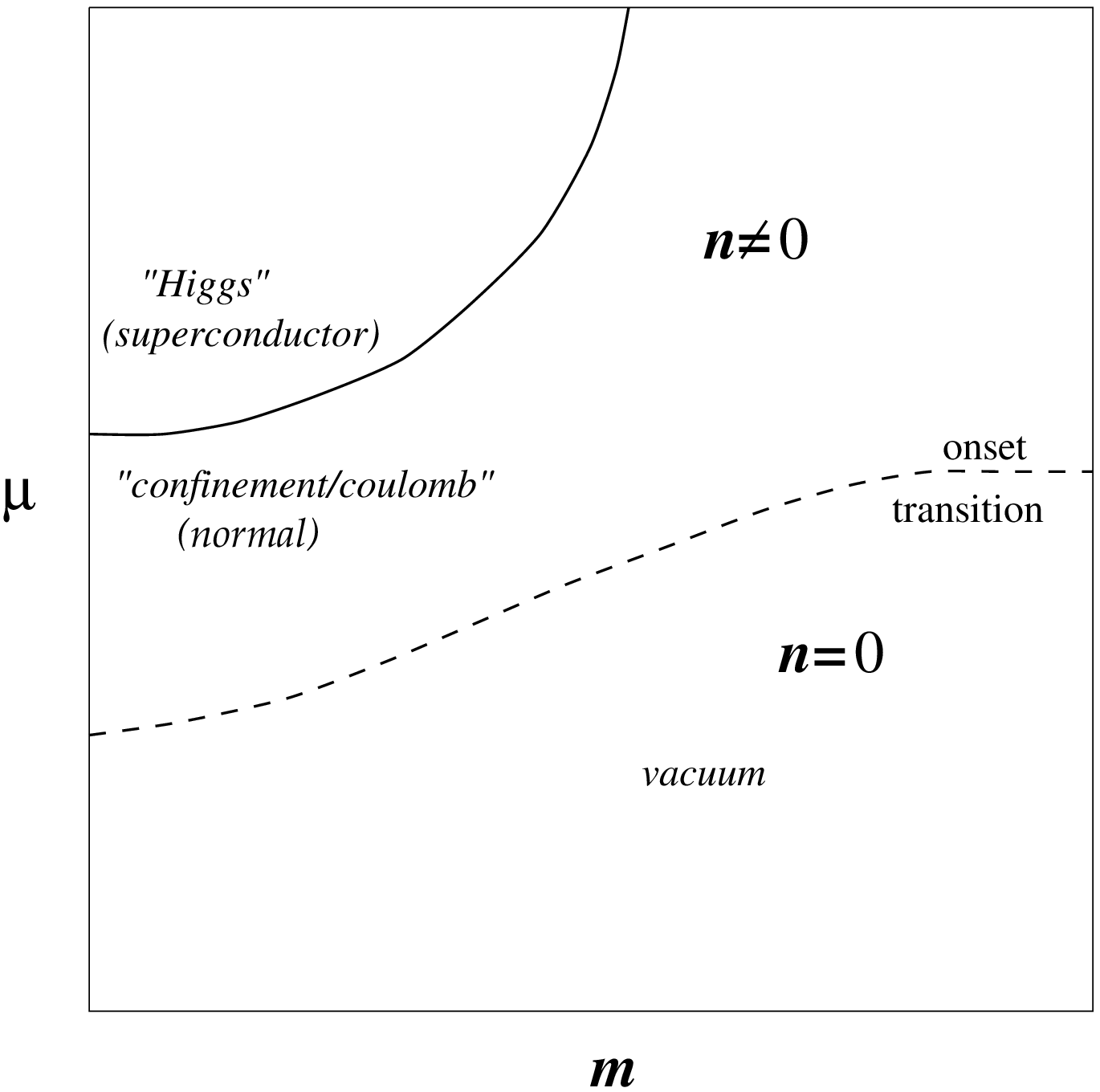}
\vskip -0.3cm
\caption{\hskip 4.1cm Figure 10.}
\vskip -0.3cm
\setcounter{figure}{10}
\end{figure}
A more compelling possibility is condensation of a gauge non-invariant
operator
\begin{equation}
qq_{\bf3}^{\prime\prime}=\chi^{tr}(x)\alpha_a t_a\chi(x),
\end{equation}
where the $t_a$ are generators of the {\bf3}, and are 
antisymmetric in the representation in which the $U_\mu$ are real.
Because $qq_{\bf3}^{\prime\prime}$ is not gauge singlet,
this results in a breaking of the SU(2) color group to U(1) 
by the Higgs mechanism; 
in other words, this is a superconducting
solution. A possible phase diagram is shown in Fig. 10. Note that since
$qq_{\bf3}^{\prime\prime}$ acts as an adjoint Higgs field, there is still a
separation between normal and superconducting phases,\cite{FS} the 
latter characterised by a massless photon.
By contrast, for QCD in the color-flavor locked state, 
continuity between high and low
density phases has been postulated.\cite{SW}

We are currently beginning to study the adjoint model using both 
HMC and multi-bosonic methods.\cite{HMMO} 
There are two aspects
of the simulation to give concern. Firstly, the HMC method permits 
a minimum $N_f=4$, which destroys asymptotic freedom for adjoint quarks, 
although that may have little direct bearing on the issue of diquark
condensation. Secondly, both algorithms use a power of $\mbox{det}M^{tr}M$
which is guaranteed positive, whereas the true measure
$\mbox{det}M$ is real but not positive definite. In effect, therefore, the
simulations incorporate an extra flavor, which allows the possibility of 
a flavor non-singlet superfluid scalar condensate 
$qq_{\bf3}^{\prime\prime\prime}=
\chi_i^{tr}\varepsilon_{ij}\chi_j$. This condensate breaks
$\mbox{U}(4)\to\mbox{Sp}(4)$ yielding six Goldstones, some of which must be
diquark states. Therefore we expect an early onset, and a phase diagram
resembling Fig. 8. These considerations point to the intriguing possibility
of a link between the sign problem generic to simulations at $\mu\not=0$ and 
a superconducting ground state.

\section*{Acknowledgments}
It is a pleasure to thank our collaborators Ian Barbour, John Kogut,
Maria-Paola Lombardo, Biagio Lucini, Istv\'an Montvay, Manfred Oevers 
and Don Sinclair. We have also greatly enjoyed discussions
with Mark Alford, Krishna Rajagopal, 
Misha Stephanov,
Jac Verbaarschot and Frank Wilczek.
This work was supported by the TMR network ``Finite temperature phase
transitions in particle physics'' EU contract ERBFMRX-CT97-0122.


\end{document}